\documentclass[11pt]{article}
\usepackage{amstex}
\usepackage{amssymb}

\numberwithin{equation}{section}

\newcommand{\sss}[1]{{\scriptscriptstyle #1}}
\newcommand{\half}{\tfrac{1}{2}}
\newcommand{\fourth}{\tfrac{1}{4}}
\newcommand{\real}{{\text{I\kern-.2em R}}}
\newcommand{\Lie}{\operatorname{\pounds}}
\newcommand{\tr}{\operatorname{tr}}

\newcommand{\e}{{\mathrm{e}}}
\newcommand{\im}{{\mathrm{i}}}
\newcommand{\pfrac}{\fracwithdelims{(}{)}}

\begin{document}

\title{Thermodynamics of Dilatonic Black Holes\\ in $n$ Dimensions}
\author{Jolien D. E. Creighton%
  \thanks{E-mail: \texttt{jolien{\char'100}avatar.uwaterloo.ca}}
  \ and Robert B. Mann%
  \thanks{E-mail: \texttt{mann{\char'100}avatar.uwaterloo.ca}}
  \\[\medskipamount]
  Dept.~of Physics, University of Waterloo,\\
  Waterloo, Ontario, Canada.  N2L~3G1}
\date{2 November, 1995}
\maketitle

\begin{center}
  arch-ive/9511012
\end{center}

\begin{abstract}\noindent
We present a formalism for studying the thermodynamics of black holes
in dilaton gravity.  The thermodynamic variables are defined on a
quasilocal surface surrounding the black hole system and are obtained
from a general class of Lagrangians involving a dilaton.  The
formalism thus accommodates a large number of possible theories and
black hole spacetimes.  Many of the thermodynamic quantities are
identified from the contribution of the action on the quasilocal
boundary.  The entropy is found using path integral techniques, and a
first law of thermodynamics is obtained.  As an illustration, we
calculate the thermodynamic quantities for two black hole solutions in
$(1+1)$ dimensions: one obtained from a string inspired theory and the
other being a Liouville black hole in the ``$R=\kappa T$'' theory with
a Liouville field.
\end{abstract}

\section{Introduction}
\label{sec:intro}

Although there is no fully viable theory of quantum gravity, we
already know many features that such a theory must possess.  If
general relativity is a good large-scale classical limit, we can
expect the existence of black holes that will radiate and eventually
evaporate completely.  An understanding of the nature of the final
stages of such evaporation remains elusive.  We do not yet know if the
information contained within the black hole will re-emerge or be
forever lost.  The resolution of this information-loss paradox will
give us much information about the nature of the quantum theory of
gravitation that we seek.

One of the major obstacles in developing a quantum theory of
gravitation is the dimensionality of our spacetime.  In
lower-dimensional spacetimes, there are many proposed theories within
which the techniques of solving simple problems are improving.  One of
the objectives in studying lower-dimensional theories is to consider
problems that are classically similar to those we have in our
four-dimensional spacetime to see how the quantum theory may be
incorporated in the lower-dimensional case.  For example, a resolution
to the information loss paradox of black hole evaporation in two
dimensions using a two-dimensional theory of quantum gravity would
most likely suggest a similar resolution in four dimensions exists.
The strength of the analogy will be increased if the resolution is
robust for many two-dimensional theories.

The first step in achieving this objective is to identify
qualitatively similar classical problems in lower dimensions to those
found in four dimensions as well as a consistent basis of comparison.
Various two-dimensional black holes are known for different
two-dimensional theories already. One of the purposes of this paper is
to give a formalism by which the thermodynamics of these systems can
be compared to the thermodynamics of the well known four-dimensional
black holes.  This is an important basis for comparison because it is
from thermodynamic considerations that we conclude that
four-dimensional black holes must evaporate.

Dilaton gravity theories are of particular interest in this regard
since they emerge as the low-energy effective field theory limit of
string theories. They are commonly studied in $(1+1)$ dimensions
because the two-dimensional analog of the Lagrangian of general
relativity yields trivial field equations.  In this paper, we consider
a large class of theories that are derivable from a Lagrangian with a
dilaton field in $n$ dimensions.  As an illustration, we allow a
possible coupling to a Maxwell field as well.  The class of
Lagrangians that we choose includes the class analyzed by
Louis-Martinez and Kunstatter~\cite{LK95} and this work generalizes
many of the results of an earlier paper~\cite{CM95} where we
restricted our analysis to four-dimensional spacetimes.

A second important feature of our analysis is the use of quasilocal
quantities to define our thermodynamic variables.  There are a number
of advantages to using quasilocal methods rather than the more usual
practice of assuming some sort of asymptotic fall-off behaviour (such
as asymptotic flatness).  Firstly, the quasilocal method is more
robust in that it can handle spacetimes of widely varying asymptotic
behaviour on the same footing.  Secondly, it provides a natural notion
for the division of a thermodynamic ``universe'' into a system and a
reservoir.  Thirdly, it is known that the statistical mechanics of
many systems of interest are well defined only for finite-sized
systems.

There are two principle influences on the methodology we use.  The
first is the quasilocal method of Brown and York~\cite{BY93a,BY93b},
who used the contribution of the action of general relativity on the
quasilocal boundary to identify thermodynamic quantities, such as
energy and angular momentum, appropriate for the spacetime contained
within the quasilocal boundary.  The particular choice of the boundary
terms of the action determine the statistical ensemble when path
integral techniques for statistical mechanics are used.  Using the
microcanonical ensemble and a ``zeroth order'' expansion of the path
integral, Brown and York recover the entropy and the first law of
thermodynamics for black hole spacetimes.

The second influence is the work of Iyer and
Wald~\cite{W93,IW94,IW95}, who identify the entropy of a black hole
spacetime with the Noether charge associated with diffeomorphism
covariance of normalized Killing vectors on the bifurcation surface.
This technique is valid for a wide range of theories described by a
very general class of diffeomorphism-covariant Lagrangians.  In fact,
within their common domain of applicability, it has been shown that
the method of Brown and York is equivalent to that of Iyer and
Wald~\cite{IW95}.

We first discuss the boundary terms arising from a general
$n$-dimensional dilaton action in section~\ref{sec:action}, and we
derive expressions for the thermodynamic variables on the quasilocal
boundary in section~\ref{sec:thermo vars}.  We also analyze a Maxwell
field with possible couplings to the dilaton in these sections as an
illustration of how matter is incorporated.  In section~\ref{sec:2D
thermo}, we restrict ourselves to the case of a $(1+1)$-dimensional
spacetime ($n=2)$ and obtain explicit expressions for the
thermodynamic quantities in Schwarzschild-like coordinates.  We apply
our definitions to two black hole spacetimes arising from different
two-dimensional theories in section~\ref{sec:examples}.  These provide
a useful check that the thermodynamic variables obtained from our
statistical mechanics approach agree with those found from
thermodynamics given an internal energy function and the first law of
thermodynamics.  Finally, some comments are made about the evaporation
of these sample black holes in section~\ref{sec:conclusions}.

\paragraph{Notation}
Throughout this paper, we use the conventions of Wald~\cite{W84}.  We
take the speed of light, the rationalized Planck constant~$\hbar$, and
Boltzmann's constant to be unity.

\section{The Dilaton Action and Boundary Terms}
\label{sec:action}

Let spacetime be an $n$-dimensional manifold on which a metric,
$g_{ab}$, is defined.  We consider a region, $\mathcal{M}$, of the
manifold that has the topology of the direct product of space-like
hypersurface, $\varSigma$, with a real (time-like) interval.

In two dimensions, the Riemann tensor has only one independent
component and can be written directly in terms of the curvature scalar
and the metric as~$R_{abcd}=Rg_{a[c}g_{d]b}$.  The Ricci tensor is
then found to be~$R_{ab}=\half g_{ab}R$, so the Einstein tensor,
$G_{ab}=R_{ab}-\half g_{ab}R$, is forced to vanish identically.  Thus,
a viable theory of gravity in two dimensions must be different from a
lower dimension version of General Relativity. We do not yet restrict
our analysis to two dimensions, but the above motivation leads us to
consider a large class of possible theories involving a dilaton field
given by the action
\begin{equation}
  I=\int_{\mathcal{M}}(\mathcal{L}_\sss{\text{D}}
  +\mathcal{L}_\sss{\text{H}}+\mathcal{L}_\sss{\text{V}}
  +\mathcal{L}_\sss{\text{M}})\,d^n\!x, \label{action}
\end{equation}
where
\begin{align*}
  \mathcal{L}_\sss{\text{D}}&=\sqrt{-g}D(\varPsi)R,\\
  \mathcal{L}_\sss{\text{H}}&=\sqrt{-g}H(\varPsi)
  g^{ab}\nabla_a\Psi\nabla_b\Psi,\\
  \mathcal{L}_\sss{\text{V}}&=\sqrt{-g}V(\varPsi),\\ \intertext{and}
  \mathcal{L}_\sss{\text{M}}&=\sqrt{-g}
  F(\varPsi,\varPhi_\sss{\text{M}};g^{ab}).
\end{align*}
The dilaton field is labeled by~$\varPsi$,
while~$\varPhi_\sss{\text{M}}$ refers to the matter fields present.
The function~$D(\varPsi)$ represents the coupling of the dilaton with
the curvature, while~$\mathcal{L}_\sss{\text{H}}$
and~$\mathcal{L}_\sss{\text{V}}$ represent the kinetic and potential
energies of the dilaton respectively.  The last term,
$\mathcal{L}_\sss{\text{M}}$, involves the matter fields as well as
possible couplings to the metric and the dilaton.  We restrict the
function~$F(\varPsi,\varPhi_\sss{\text{M}};g^{ab})$ by requiring that
it contains no derivatives of the metric or the dilaton.  This
restriction guarantees that the matter will not contribute to the
boundary terms of interest. Furthermore, we require that the
functions~$D(\varPsi)$, $H(\varPsi)$, and~$V(\varPsi)$ contain no
derivatives of the dilaton field.

\subsection{Variational Boundary Terms}
Field equations are those that result in the vanishing of the
variation of the action when certain conditions on the boundary of the
region~$\mathcal{M}$ are imposed.  The exact nature of the boundary
conditions are often ignored and interest is placed on the field
equations alone.  However, since we consider only a \emph{finite}
region, $\mathcal{M}$, the exact nature of the boundary conditions
becomes important.  Thus, we carefully examine the variation of the
\emph{gravitational} sector of the theory, that is, the Lagrangian
density~$\mathcal{L}_\sss{\text{G}}=\mathcal{L}_\sss{\text{D}}
+\mathcal{L}_\sss{\text{H}}+\mathcal{L}_\sss{\text{V}}$.  Our analysis
is similar to that of Burnett and Wald~\cite{BW90}, but our Lagrangian
density is more general.

We introduce a one-parameter family of variations of the spacetime
metric and the dilaton field.  We find the resulting variation in the
Lagrangian density of the gravitational sector is:
\begin{gather}
  \mathcal{L}_\sss{\text{G}} = \sqrt{-g}\,\bigl( (E_g)_{ab} \delta
  g^{ab} + (E_\varPsi) \delta \varPsi + \nabla_a \rho^a \bigr),
  \label{var grav action} \\ \intertext{where} \begin{split}
  (E_g)_{ab} &= D(\varPsi)G_{ab} + g_{ab}\nabla^2 D(\varPsi) -
  \nabla_a\nabla_b D(\varPsi)\\ &\quad + H(\varPsi) \bigl(
  \nabla_a\varPsi\nabla_b\varPsi - \half g_{ab}(\nabla\varPsi)^2
  \bigr) - \half g_{ab} V(\varPsi),\\ \end{split} \label{metric EOM}\\
  (E_\varPsi) = \frac{dD}{d\varPsi}\,R - \frac{dH}{d\varPsi}\,
  (\nabla\varPsi)^2 - 2H(\varPsi)\nabla^2\varPsi +
  \frac{dV}{d\varPsi}, \label{dilaton EOM} \\
  \intertext{and~$\rho^a=\rho^a(g,\varPsi,\delta g,\delta\varPsi)$ is}
  \begin{split} \rho^a &= D(\varPsi) \bigl( \nabla^a(g_{cd}\delta
  g^{cd}) - \nabla_b \delta g^{ab} \bigr) - g_{cd}\delta
  g^{cd}\nabla^a D(\varPsi) + \delta g^{ab}\nabla_b D(\varPsi)\\
  &\quad + (2H(\varPsi)\nabla^a\varPsi) \delta \varPsi.\\
  \end{split}\label{boundary}
\end{gather}
The variational boundary terms are all contained in $\rho^a$.  If the
variation satisfies sufficient conditions such that there is no
boundary contribution, then the field equations are given
by~$(E_g)_{ab}=\half T_{ab}$ and~$(E_\varPsi)=\half U$, where $T_{ab}$
is the stress tensor arising from the coupling between the matter
fields and the spacetime metric, and $U$ is an analogous quantity
arising from couplings between the matter fields and the dilaton.
These source quantities are defined as
\begin{gather}
  T^{ab}=\frac{2}{\sqrt{-g}}\,\frac{\delta I_\sss{\text{M}}}{\delta
  g_{ab}}\label{SEM source}\\ \intertext{and}
  U=\frac{2}{\sqrt{-g}}\,\frac{\delta I_\sss{\text{M}}}{\delta
  \varPsi}.\label{dilaton source}
\end{gather}
It is an interesting property that the two equations of motion given
in equations~\eqref{metric EOM} and~\eqref{dilaton EOM} are related
by~$\nabla^a(E_g)_{ab}=-\half(E_\varPsi)\nabla_b\varPsi$.  An
immediate consequence is that, if the matter fields do not couple to
the dilaton (so that~$U=0$), and if the second equation of motion,
$(E_\varPsi)=0$, is satisfied, then the quantity~$(E_g)_{ab}$ is
divergenceless.  If we demand that the first equation of motion,
$(E_g)_{ab}=\half T_{ab}$, is satisfied, then it follows that the
stress tensor of the matter must also be divergenceless.

Let us consider more closely the contribution of the boundary terms.
Suppose an element of the boundary has a normal vector~$n^a$, oriented
outwards for space-like normal vectors and inwards for time-like
vectors, and normalized so that~$n^an_a=\iota$ with~$\iota=\pm1$ for
space-like/time-like~$n^a$.  First and second fundamental forms on the
boundary (the induced metric on the boundary and the extrinsic
curvature of the boundary) can be constructed; they are
$\gamma_{ab}=g_{ab}-\iota n_a n_b$
and~$\varTheta_{ab}=-\half\Lie_n\gamma_{ab}$ respectively.  We assume
that the boundary is fixed under the variations so that variations of
the normal dual-vector on the boundary are proportional to the normal
dual-vector.  Therefore, on the boundary element, we find $\delta
g^{ab}=2\iota n^{(a}\delta n^{b)}+\delta\gamma^{ab}$, with
$\gamma^{ab}\delta n_b=0$ and~$\delta\gamma^{ab}n_b=0$.

The contribution of the boundary terms to the variation of the action
functional of equation~\eqref{action} is related to~$n_a\rho^a$.  One
can show that
\begin{align}
  \sqrt{|\gamma|}\, n_a\rho^a &= \pi^{ab} \delta\gamma_{ab} + \varPi
  \delta\varPsi + \delta\alpha + \sqrt{|\gamma|}\,\mathcal{D}_a
  \beta^a, \label{norm bnd}\\ \intertext{where} \pi^{ab} &=
  \sqrt{|\gamma|}\, \Bigl( \gamma^{ab} n^c\nabla_c D(\varPsi) +
  D(\varPsi) \bigl( \varTheta^{ab} - \gamma^{ab} \tr(\varTheta) \bigr)
  \Bigr), \label{momentum conj metric} \\ \varPi &= -2
  \sqrt{|\gamma|}\, \Bigl( \tr(\varTheta)\, \frac{dD}{d\varPsi} -
  H(\varPsi) n^c\nabla_c\varPsi \Bigr), \label{momentum conj dil} \\
  \alpha &= 2\sqrt{|\gamma|}\, D(\varPsi) \tr(\varTheta), \label{bnd
  term one} \\ \beta^a &= D(\varPsi) \gamma^a_c \delta n^c, \label{bnd
  term two}
\end{align}
and $\mathcal{D}_a$ is the derivative operator compatible with the
induced metric $\gamma_{ab}$.  The volume element on the boundary
is~$(-\iota\det(\gamma))^{1/2}$ which we write simply
as~$\sqrt{|\gamma|}$.  The last term in equation~\eqref{norm bnd} is a
total derivative; the integral of this term is proportional to the
projection of $\beta^a$ onto the normal of the boundary of the
boundary element.  We assume that this normal is orthogonal to~$n^a$,
and thus it is orthogonal to $\beta^a$ as well.  Thus, we can ignore
the last term in equation~\eqref{norm bnd}.

\subsection{Noether Currents and Charges}

The Lagrangian density of the gravitational sector is covariant under
diffeomorphisms.  Wald \cite{W93} has shown that there is a conserved
Noether current associated with this covariance.  The conservation of
this Noether current implies the existence of a Noether charge, which
can be used to find the entropy of black hole spacetimes.

Consider a variation of the Lagrangian density of the gravitational
sector, $\mathcal{L}_\sss{\text{G}}$, that is associated with
diffeomorphisms along some vector $\xi^a$.  The field variations are
given by
\begin{gather}
  \delta g_{ab} = \Lie_\xi g_{ab} = 2 \nabla_{(a}\xi_{b)} \label{delta
  g}\\ \intertext{and} \delta\varPsi = \Lie_\xi\varPsi =
  \xi^a\nabla_a\varPsi.  \label{delta Psi}
\end{gather}
Because of the diffeomorphism covariance of the Lagrangian density,
\begin{equation}
  \delta\mathcal{L}_\sss{\text{G}} = \Lie_\xi
  \mathcal{L}_\sss{\text{G}}= \nabla_a(\xi^a
  \mathcal{L}_\sss{\text{G}}), \label{delta L grav}
\end{equation}
the quantity
\begin{equation}
  \sqrt{-g}J^a[\xi] = \sqrt{-g}\rho^a(g,\varPsi,\Lie_\xi g,\Lie_\xi
  \varPsi) - \xi^a\mathcal{L}_\sss{\text{G}} \label{Noether current}
\end{equation}
is divergenceless when the equations of motion ($(E_g)_{ab}=0$
and~$(E_\varPsi)=0$) hold.

Using the expression for $\rho^a$ from equation~\eqref{boundary}, as
well as the expressions for the diffeomorphic variations of the fields
in equations~\eqref{delta g} and~\eqref{delta Psi}, we find an
explicit form for the Noether current:
\begin{equation}
  J^a[\xi] = -2\nabla_b \bigl( 2\xi^{[a}\nabla^{b]} D(\varPsi) +
  D(\varPsi) \nabla^{[a}\xi^{b]}\bigr) + 2\xi_b (E_g)^{ab}
  \label{Noether current II}
\end{equation}
with $(E_g)^{ab}$ given in equation~\eqref{metric EOM}.  When the
(sourceless) equation of motion, $(E_g)^{ab}=0$, holds, the Noether
current is divergenceless and we can define a charge density,
$Q[\xi]$, on a closed space-like $(n-2)$-dimensional hypersurface
\begin{equation}
  J^a[\xi] = n^{ab}\nabla_b Q[\xi], \label{Noether charge density}
\end{equation}
where $n^{ab}$ is the normal bi-vector to the $(n-2)$~surface.  The
normal bi-vector has an orientation such that, if $R^a$ is a radially
outward-directed space-like vector and $T^a$ is a future-directed
time-like vector, then~$n^{ab}T_aR_b<0$.  We find that the Noether
charge density is
\begin{equation}
  Q[\xi]=n^{ab}(2\xi_a\nabla_b D(\varPsi) + D(\varPsi)\nabla_a\xi_b).
  \label{on shell Noether charge density}
\end{equation}
The integral of this quantity over the $(n-2)$~surface is the
conserved Noether charge.

\subsection{Boundary Terms from a Dilaton Maxwell Action}

Let us illustrate the effect of additional fields on the boundary
terms we have constructed.  As an example, consider the following
Lagrangian density
\begin{equation}
  \mathcal{L}_\sss{\text{M}} = \sqrt{-g}\fourth W(\varPsi) {\frak
  F}^{ab}{\frak F}_{ab}, \label{Maxwell Lagrangian}
\end{equation}
where ${\frak F}_{ab}=2\nabla_{[a}{\frak A}_{b]}$ is the Maxwell field
strength and ${\frak A}_a$ the potential.  The Maxwell field is
coupled to both the metric and, through the function~$W(\varPsi)$, the
dilaton.  We assume that $W(\varPsi)$ is a function of the dilaton
alone, and that it does not contain any derivatives of the dilaton.
Under variations of the metric, the dilaton, and the Maxwell field
potential, the induced variation of the dilaton Maxwell Lagrangian is
\begin{align}
  \delta\mathcal{L}_\sss{\text{M}} &= \bigl( -\half T_{ab} \delta
  g^{ab} - \half U \delta \varPsi + (E_{\frak A})^a \delta {\frak A}_a
  + \nabla_a \varrho^a \bigr) \label{var Maxwell Lagrangian}\\
  \intertext{with} T_{ab} &= -W(\varPsi)({\frak F}_{ac}{\frak F}_b{}^c
  - \fourth g_{ab}{\frak F}^{cd}{\frak F}_{cd}), \label{Maxwell SEM
  source}\\ U &= -\frac{1}{2}\,\frac{dW}{d\varPsi}\,{\frak F}^{ab}
  {\frak F}_{ab},\label{Maxwell dilaton source}\\ (E_{\frak A})^a &=
  \nabla_b \bigl( W(\varPsi){\frak F}^{ab} \bigr), \label{Maxwell
  EOM}\\ \intertext{and} \varrho^a &= W(\varPsi) {\frak F}^{ab} \delta
  {\frak A}_b.  \label{Maxwell bnd}
\end{align}
The sourceless dilaton Maxwell equations are~$(E_{\frak A})^a=0$.

Let ${\frak U}_a=\gamma^b_a{\frak A}_b$ be the projection of the
Maxwell field potential onto the boundary of~$\mathcal{M}$.  One finds
that the projection of the variation of the Maxwell field potential is
just the variation of the projection of the Maxwell field potential.
Thus, on the boundary~$\partial\mathcal{M}$, we find
that~$\sqrt{|\gamma|}n_a\varrho^a=\varpi^a\delta{\frak U}_a$ with
\begin{equation}
  \varpi^a=-\sqrt{|\gamma|}\,W(\varPsi) {\frak F}^{ab} n_b
  \label{Maxwell momentum}
\end{equation}
being the momentum conjugate to the boundary Maxwell field potential.

Because of the extra term in the total Lagrangian density and the
additional boundary contribution, $\varrho^a$, there is an extra
contribution to the Noether current
\begin{equation}
  \sqrt{-g}(J_{\text{extra}})^a[\xi] = \sqrt{-g}\,\varrho^a[\xi] -
  \xi^a \mathcal{L}_\sss{\text{M}}, \label{extra Noether current}
\end{equation}
where the variations are diffeomorphism induced
\begin{equation}
  \delta {\frak A}_b = \Lie_\xi{\frak A}_b = \xi^a\nabla_a{\frak A}_b
  + {\frak A}_a\nabla_b\xi^a.  \label{delta A}
\end{equation}
When this extra piece is included in equation~\eqref{Noether current
II}, we find that
\begin{equation}
  \begin{split} J^a &= \nabla_b \bigl( W(\varPsi){\frak F}^{ab}{\frak
    A}_c\xi^c - 4\xi^{[a}\nabla^{b]}D(\varPsi) -
    2D(\varPsi)\nabla^{[a}\xi^{b]} \bigr) \\ &\qquad + \xi_b \bigl(
    2(E_g)^{ab} - T^{ab} \bigr) - \xi^b{\frak A}_b (E_{\frak A})^a.\\
    \end{split} \label{Maxwell Noether current}
\end{equation}
If the equations of motion, $(E_g)^{ab}=\half T^{ab}$ and $(E_{\frak
A})^a=0$, hold, the Noether current becomes divergenceless, so a
Noether charge may be constructed.  This Noether charge is given by
equation~\eqref{on shell Noether charge density} with an additional
term
\begin{equation}
  Q_{\text{extra}} = W(\varPsi)(n^a{\frak E}_a)(\xi^b{\frak A}_b).
  \label{extra Noether charge}
\end{equation}
Here, ${\frak E}_a$~is the electric field with~$n^{ab}{\frak
F}_{ab}=-2n^a{\frak E}_a$.

\section{Thermodynamic Variables}
\label{sec:thermo vars}

As in the previous section, we consider a region of the spacetime
manifold, $\mathcal{M}$, which has the topology of the direct product
of a space-like hypersurface, $\varSigma$, with a time-like interval.
Suppose that this interval is parameterized with the parameter~t; the
associated vector, $t^a$, satisfies~$t^a\nabla_a t=1$.  The space-like
leaves of the foliation are $\varSigma_t$.  Our primary interest shall
be restricted to the outer time-like boundary, $\mathcal{T}$, of the
region, $\mathcal{M}$.  By~$n^a$, $\gamma_{ab}$, and~$\varTheta_{ab}$,
we shall refer to the normal, the induced metric, and the extrinsic
curvature of $\mathcal{T}$ respectively.  We shall write~$u^a$,
$h_{ab}$, and~$K_{ab}$ respectively for the same quantities on the
leaves of the foliation.  Furthermore, the lapse function and shift
vector are defined by~$N=-t^au_a$ and~$N^a=h^a_bt^b$ respectively.  We
demand that~$u^a$ and~$n^b$ be orthogonal everywhere on~$\mathcal{T}$.
Moreover, the foliation of~$\mathcal{M}$ induces a foliation
of~$\mathcal{T}$ into $(n-2)$~surfaces, $\mathcal{B}_t$.  These closed
$(n-2)$~surfaces define the boundary of the gravitating system; we
shall call this boundary the \emph{quasilocal boundary}, or,
equivalently, the \emph{system boundary} for thermodynamics.  The
specification of this boundary is not gauge invariant in that,
given~$\mathcal{T}$, there are many possible choices of the parameter
of foliation.  Physically, we think of the quasilocal boundary as
being defined by a class of observers whose world lines form a
congruence on~$\mathcal{T}$ with tangent vectors~$u^a$.  Due to
hypersurface orthogonality, these observers are zero vorticity
observers.  The acceleration of these observers is given
by~$a^b=u^a\nabla_au^b=N^{-1}h^{ab}\nabla_aN$.  Finally, the metric on
the quasilocal boundary is given by~$\sigma_{ab}=\gamma_{ab}+u_au_b$,
and, if the quasilocal boundary is viewed as the outer boundary,
$(\partial\varSigma_t)_{\text{out}}$, then it has an extrinsic
curvature~$k_{ab}=-\half\Lie_n\sigma_{ab}$.

\subsection{Quasilocal Quantities}

As we have shown in section~\ref{sec:action}, the generation of field
equations from the action depends upon the choice of appropriate
boundary conditions.  A natural boundary condition is the fixation of
all the fields on the boundaries (whereas the derivatives of the field
variations are not so constrained).  Although these conditions are not
appropriate for the action given by equation~\eqref{action}, we see
from equations~\eqref{norm bnd} and~\eqref{bnd term one} that the
action
\begin{equation}
  I^1 = I - 2\int_\varSigma \sqrt{h}\,D(\varPsi)\tr(K) d^{(n-1)}\!x -
  2\int_{\mathcal{T}} \sqrt{-\gamma}\,D(\varPsi)\tr(\varTheta)
  d^{(n-1)}\!x - I^0 \label{action1}
\end{equation}
\emph{is} appropriate.  In the above, we have included boundary
integrals for the initial space-like boundary and the outer time-like
boundary only; similar terms could be added for additional boundary
elements.  An arbitrary functional that is linear in the boundary
fields, $I^0$, is allowed.  We view the effect of this quantity as a
specification of a reference spacetime that effectively defines the
zero of the energy.  The quantity~$I$ is just the action given in
equation~\eqref{action}.

We obtain the momenta conjugate to the field configurations on the
boundaries by taking functional derivatives of the action~$I^1$,
evaluated ``on shell'' (i.e., when the field equations are satisfied),
with respect to the boundary field configuration.  Thus, the momenta
conjugate to the boundary metrics are~$p^{ab}-p^{ab}_\sss0=-(\delta
I^1/\delta h_{ab})_{c\ell}$ for the space-like boundary
and~$\pi^{ab}-\pi^{ab}_\sss0=(\delta I^1/\delta\gamma_{ab})_{c\ell}$
for the time-like boundary~$\mathcal{T}$.  Here, the
subscripted~``$c\ell$'' indicates evaluation on a classical solution.
These momenta are given explicitly as
\begin{align}
  p^{ab} &= -\sqrt{h} \Bigl( h^{ab}u^c\nabla_c D(\varPsi) + D(\varPsi)
  \bigl( K^{ab} - h^{ab}\tr(K) \bigr)\Bigr) \label{ini geom
  momentum}\\ \intertext{and} \pi^{ab} &= \sqrt{-\gamma} \Bigl(
  \gamma^{ab}n^c\nabla_c D(\varPsi) + D(\varPsi) \bigl( \varTheta^{ab}
  - \gamma^{ab}\tr(\varTheta) \bigr)\Bigr), \label{bnd geom momentum}
\end{align}
where the terms~$p^{ab}_\sss0$ and~$\pi^{ab}_\sss0$ are associated
with the boundary variations of the functional~$I^0$.  Similar momenta
conjugate to the matter and dilaton fields can be obtained.  The
latter are~$P-P_\sss0=-(\delta I^1/\delta\varPsi)_{c\ell}$ for the
space-like boundary and~$\varPi-\varPi_\sss0=(\delta
I^1/\delta\varPsi)_{c\ell}$ for the time-like boundary with
\begin{align}
  P &= -\sqrt{h} \Bigl( 2D(\varPsi)u^a\nabla_a\varPsi
  -2\frac{dD}{d\varPsi}\,\tr(K) \Bigr) \label{ini dil momentum} \\
  \intertext{and} \varPi &= \sqrt{-\gamma} \Bigl(
  2D(\varPsi)n^a\nabla_a\varPsi -2\frac{dD}{d\varPsi}\,\tr(\varTheta)
  \Bigr).  \label{bnd dil momentum}
\end{align}

The momenta on the~$\mathcal{T}$ boundary contain useful information
about the gravitating system enclosed, and we use this information to
construct our quasilocal quantities.  For a given a quasilocal
surface, the variation boundary metric~$\gamma_{ab}$ can be decomposed
into its projections normal and tangential to the surface:
\begin{equation}
  \delta\gamma_{ab} = \sigma^c_a \sigma^d_b \delta \sigma_{cd} -
  \frac{2}{N}\,u_a u_b \delta N - \frac{2}{N}\,u_{(a}\sigma_{b)c}
  \delta N^c.  \label{decomp bnd metric}
\end{equation}
Let us introduce the following densities on the quasilocal surface:
\begin{align*}
  \mathcal{E}&=\frac{2}{N}\,u_au_b(\pi^{ab}-\pi^{ab}_\sss0), \\
  \mathcal{J}^c&=-\frac{2}{N}\,u_a\sigma^c_b(\pi^{ab}-\pi^{ab}_\sss0),
  \\ \intertext{and}
  \mathcal{S}^{cd}&=\frac{2}{N}\,\sigma^c_a\sigma^d_b(\pi^{ab}
  -\pi^{ab}_\sss0).
\end{align*}
These are the quasilocal surface energy density, momentum density, and
stress density; they are just the corresponding projections of the
momenta conjugate to the boundary metric.  We can evaluate these
quantities using equation~\eqref{bnd geom momentum} and the
relationship~\cite{BY93a}
\begin{equation}
  \varTheta_{ab} = k_{ab} + u_a u_b n^c a_c + 2u_{(a} \sigma^c_{b)}
  n^d K_{cd}.  \label{decomp Theta}
\end{equation}
We find
\begin{gather}
  \mathcal{E} = -2 \sqrt{\sigma} \bigl( n^a\nabla_a D(\varPsi) -
  D(\varPsi) \tr(k) \bigr) - \mathcal{E}_\sss0, \label{energy
  density}\\ \begin{split} \mathcal{J}^c &= 2\sqrt{\sigma} D(\varPsi)
  n_a \sigma^c_b K^{ab} - (\mathcal{J}_\sss0)^c \\ &=
  -\frac{2\sqrt{\sigma}}{\sqrt{h}}\,n_a \sigma^c_b p^{ab} -
  (\mathcal{J}_\sss0)^c, \\ \end{split}\label{momentum density}\\
  \intertext{and} \mathcal{S}^{cd} = 2\sqrt{\sigma} \biggl(
  \sigma^{cd} n^a\nabla_a D(\varPsi) + D(\varPsi) \Bigl( k^{cd} -
  \sigma^{cd} \bigl( \tr(k) - n^a a_a \bigl) \Bigl) \biggl) -
  (\mathcal{S}_\sss0)^{cd}, \label{stress density}
\end{gather}
where $\mathcal{E}_\sss0$, $(\mathcal{J}_\sss0)^c$,
and~$(\mathcal{S}_\sss0)^{cd}$ arise from the~$\pi^{ab}_\sss0$
component of the momentum.

The dilaton field is a scalar field and does not require any
decomposition.  We define the
quantity~$\mathcal{Y}=N^{-1}(\varPi-\varPi_\sss0)$:
\begin{equation}
  \mathcal{Y} = \sqrt{\sigma} \Bigl( 2D(\varPsi) n^a\nabla_a\varPsi -
  2\frac{dD}{d\varPsi} \bigl( \tr(k) - n^a a_a \bigr) \Bigr) -
  \mathcal{Y}_\sss0 \label{dilaton pressure}
\end{equation}
as the quasilocal surface dilaton ``pressure.''  Therefore, under
arbitrary field variations of the action~$I^1$, the contribution from
the~$\mathcal{T}$ boundary (if we ignore the contribution of any
matter fields that may be present) is
\begin{equation}
  \delta I^1 |_{\mathcal{T}} = \int_{\mathcal{T}} \bigl( -\mathcal{E}
  \delta N + \mathcal{J}_a \delta N^a + N ( \half\mathcal{S}^{ab}
  \delta \sigma_{ab} + \mathcal{Y} \delta \varPsi ) \bigr)
  d^{(n-1)}\!x.  \label{var action1 bnd}
\end{equation}

Notice that $\mathcal{E}$, $\mathcal{J}_a$, $\sigma_{ab}$,
and~$\varPsi$ are all functions on the phase-space, i.e., they are
functions of the phase-space variables:
\{$(p^{ab},h_{ab})$,$(P,\varPsi)$\}.  Such quantities we shall call
\emph{extensive variables}.  However, the lapse and the shift cannot
be constructed out of the phase space variables; such quantities are
\emph{intensive variables}.  Thus, the action~$I^1$ is the appropriate
choice when the extensive quantities~$\sigma_{ab}$ and~$\varPsi$ and
the intensive quantities~$N$ and~$N^a$ are fixed on the quasilocal
surface.

Define the quasilocal energy as the integral of the quasilocal surface
energy density over the quasilocal surface:
\begin{equation}
  E=\int_{\mathcal{B}} \mathcal{E} d^{(n-2)}\!x.  \label{quasi energy}
\end{equation}
As mentioned earlier, this definition is foliation-dependent;
different classes of observers will obtain different values of the
quasilocal energy.

\subsection{Conserved Quantities}

Although the quasilocal energy is foliation-dependent, it is possible
to construct quantities on the quasilocal surfaces that are
foliation-independent.  We shall call these quantities conserved since
various classes of observers will agree on the value of the quantity.
However, the spacetime must satisfy certain conditions in order to
define a conserved quantity.  The first requirement is that the
spacetime possess a vector field, $\xi^a$, such that the Lie
derivative of all the fields along this vector field vanish.  Such a
vector field provides natural choices for the boundary~$\mathcal{T}$
as those boundaries that contain a congruence of these vectors.  (Note
that~$\xi^a$ need not be defined over the entire manifold; it just
needs to be defined on~$\mathcal{T}$.)

Consider the equation of motion~$2(E_g)_{ab}=T_{ab}$.  Project the
first component normal to~$\mathcal{T}$ and the second
onto~$\mathcal{T}$ and use the Gauss Codacci
relation~$n^a\gamma^{bc}R_{ab}=-\mathcal{D}_a\bigl(\varTheta^{ac}
-\gamma^{ac}\tr(\varTheta)\bigr)$, where $\mathcal{D}_a$ is the
derivative operator compatible with~$\gamma_{ab}$, to obtain
\begin{equation}
  2\mathcal{D}_a(\pi^{ac}-\pi^{ac}_\sss0) = \varPi \mathcal{D}^c
  \varPsi - \sqrt{-\gamma} n^a \gamma^{bc} T_{ab}.  \label{bnd geom
  momentum source}
\end{equation}
We see that the momentum conjugate to the boundary metric is not
divergenceless on the boundary: it has a source term.  However, if we
contract equation~\eqref{bnd geom momentum source} with the Killing
vector~$\xi^a$, we find that
\begin{equation}
  \mathcal{D}_a\bigl(\xi_b(\pi^{ab}-\pi^{ab}_\sss0)\bigr) =
  -\sqrt{-\gamma}\,n^a \xi^b T_{ab}.  \label{bnd geom momentum source
  II}
\end{equation}
Integrate this over the boundary~$\mathcal{T}$.  We obtain
\begin{gather}
  K(\partial\mathcal{T}_{\text{final}}) -
  K(\partial\mathcal{T}_{\text{initial}}) = \int_{\mathcal{T}}
  \sqrt{-\gamma}\,n^a \xi^b T_{ab} d^{(n-1)}\!x, \label{charge
  source}\\ \intertext{where} K(\mathcal{B}) = -\int_{\mathcal{B}}
  \frac{1}{N}\,u_a \xi_b(\pi^{ab} - \pi^{ab}_\sss0) d^{(n-2)}\!x.
  \label{conserved charge}
\end{gather}
When the right hand side of equation~\eqref{charge source} vanishes,
we see that the value of $K[\xi]$ becomes independent of the
particular cut of~$\mathcal{T}$ on which it is evaluated.  Therefore,
it represents a conserved quasilocal charge.  Thus, an additional
condition for the existence of a conserved charge is the vanishing of
the right hand side of equation~\eqref{charge source}.  This may occur
if the boundary~$\mathcal{T}$ is chosen outside of the matter
distribution (so that~$T_{ab}=0$), or if the projection~$n^a\xi^b
T_{ab}$ vanishes for the specific type of matter considered.

If $\varphi^a$ is a space-like azimuthal Killing vector, then we can
define an angular momentum, $L=K[\varphi]$.  If the quasilocal
surface, $\mathcal{B}$, is taken so that it contains the orbits of the
Killing vector, then
\begin{equation}
  L=\int_{\mathcal{B}} \mathcal{J}_a \varphi^a d^{(n-2)}\!x.
  \label{angular momentum}
\end{equation}
Alternately, when~$\xi^a$ is time-like, we can define a conserved mass
as~$M=-K[\xi]$.  For a static spacetime, $\xi^a$ is surface forming
and so we can choose an $(n-2)$~surface, $\mathcal{B}$, for which the
Killing vector is proportional to the time-like normal.  Then, the
mass can be written as
\begin{equation}
  M=\int_{\mathcal{B}} N \mathcal{E} d^{(n-2)}\!x.  \label{mass}
\end{equation}
By comparing equation~\eqref{mass} with equation~\eqref{quasi energy},
we see that the quasilocal energy is not the same as the conserved
mass.  However, note that for asymptotically flat spacetimes these two
quantities approach the same value in the asymptotic region since the
lapse approaches unity.

\subsection{Entropy of Static Black Hole Space-Times}

Associated with event horizons within a thermodynamic system is an
entropy: it is a notion of the information of the total system lost to
the observers on the system boundary.  In the absence of any matter
fields, this is the only contribution to the entropy of the
thermodynamic system.  Here we explore such contributions for the case
of systems containing a static black hole.  Once again we will ignore
possible matter fields for simplicity.

Following Brown and York~\cite{BY93b}, we derive the entropy from
Euclidean path integral techniques using a ``micro-canonical'' action.
The micro-canonical action is one for which the extensive variables
(alone) must be fixed on the quasilocal boundary.  We see from
equation~\eqref{var action1 bnd} that the action~$I^1$ is \emph{not}
the micro-canonical action because it requires the fixation of both
extensive and intensive variables on the quasilocal boundary.
According to Iyer and Wald~\cite{IW95}, we should consider the action
\begin{equation}
  I_{\text{m}} = \int_{\mathcal{M}} \mathcal{L}_{\text{G}} d^n\!x +
  \int_{\partial\mathcal{M}} \sqrt{\sigma} Q[t] d^{(n-1)}\!x,
  \label{micro action}
\end{equation}
where we are only interested in the boundary element that is the
history of the quasilocal boundary.  $Q[t]$ is just the Noether charge
associated with the diffeomorphism covariance of the Lagrangian along
the vector field, $t^a$, the time-like Killing vector of the static
spacetime.

Let us show that this action \emph{is} the micro-canonical action.
First, we need to evaluate the quantity
\begin{equation}
  Q[t] = n^{ab} \bigl( 2t_a \nabla_b D(\varPsi) + D(\varPsi) \nabla_a
  t_b \bigr).  \label{Noether Charge [t] I}
\end{equation}
The first term is just~$-2Nn^a\nabla_aD(\varPsi)$, while a calculation
of the second term
yields~$2D(\varPsi)(Nu^au^b\varTheta_{ab}+N^au^b\varTheta_{ab})$ (see
\cite{IW95}).  Now using equation~\eqref{decomp Theta} as well as the
definitions of the quasilocal energy and momentum,
equations~\eqref{energy density} and~\eqref{momentum density}, we find
\begin{equation}
  \sqrt{\sigma} Q[t] = N \mathcal{E} - N^a \mathcal{J}_a - \alpha,
  \label{Noether Charge [t] II}
\end{equation}
where~$\alpha$ is given in equation~\eqref{bnd term one}.  Therefore,
the boundary contribution to the variation of the
action~$I_{\text{m}}$ is
\begin{equation}
  \begin{split} \delta I_{\text{m}} &= \int_{\partial\mathcal{M}}
    \bigl( \delta ( \sqrt{\sigma} Q[t] ) + \sqrt{-\gamma} n_a \rho^a
    \bigr) d^{(n-1)}\!x \\ &= \int_{\partial\mathcal{M}} \bigl( \delta
    ( N \mathcal{E} - N^a \mathcal{J}_a ) + ( \sqrt{-\gamma} n_a
    \rho^a - \delta \alpha ) \bigr) d^{(n-1)}\!x \\ &=
    \int_{\partial\mathcal{M}} \bigl( N \delta \mathcal{E} - N^a
    \delta \mathcal{J}_a + N ( \half \mathcal{S}^{ab} \delta
    \sigma_{ab} + \mathcal{Y} \delta \varPsi ) \bigr) d^{(n-1)}\!x, \\
    \end{split} \label{var micro action}
\end{equation}
where the second term in the second line has been calculated in
equation~\eqref{var action1 bnd}.  Clearly, it is the extensive
variables that must be held fixed on the quasilocal boundary in order
for the variations of the action~$I_{\text{m}}$ to yield the field
equations.  Thus, $I_{\text{m}}$ is a micro-canonical action.

The prescription of Brown and York~\cite{BY93b} for finding the
entropy is the following.  First, we make the Wick rotation
$t\to\tau=\im t$.  The extensive variables are invariant under this
transformation, while the intensive variables become complex.  The
initial and final space-like hypersurfaces are identified.  Because
the system contains an event horizon, the Euclidean manifold is
degenerate at this ``point,'' and the black hole interior is removed
from the Euclidean manifold.  In order to prevent a conical
singularity from forming at the event horizon, the period of the
identification is set to~$\Delta\tau=2\pi/\varkappa_\sss{\text{H}}$,
where~$\varkappa_\sss{\text{H}}=h^{ab}(\partial_aN)(\partial_bN)$
(evaluated on the event horizon) is the surface gravity.  The
micro-canonical density matrix, $\nu$, is formally given by
\begin{equation}
  \begin{split} \nu\mbox{[extensive variables]} &= \int\text{[periodic
    histories]} \,\e^{{\bar I}_{\text{m}}{\text{[extensive
    variables]}}} \\ &\approx \e^{ {\bar I}_{\text{m}} |_{c\ell} }
    \qquad \text{(zeroth order)},\\ \end{split} \label{micro density
    matrix}
\end{equation}
where~${\bar I}_{\text{m}}$ is the Euclideanized micro-canonical
action which is just the usual micro-canonical action calculated on
the Euclidean manifold, $\bar{\mathcal{M}}$, and the
subscripted~``$c\ell$'' indicates evaluation on the analytic
continuation of the classical solution.  The second line gives the
evaluation of the path integral to the zeroth order of quantum
corrections.  The entropy, $S$, is just the logarithm of the
micro-canonical density matrix, $S\approx{\bar I}_{\text{m}}$.  Thus,
\begin{align*}
  S &\approx \Biggl[ \int_{\bar{\mathcal{M}}}
  \mathcal{L}_\sss{\text{G}} d^n\!{\bar x} +
  \int_{\partial{\bar{\mathcal{M}}}} \sqrt{\sigma} Q[t]
  d^{(n-1)}\!{\bar x} \Biggr]_{c\ell} \\ &= \int d\tau \Biggl[
  -\int_\varSigma u_a ( N^{-1} t^a \mathcal{L}_\sss{\text{G}} )
  d^{(n-1)}x + \int_{\mathcal{B}} \sqrt{\sigma} Q[t] d^{(n-2)}x
  \Biggr]_{c\ell},
\end{align*}
where~$\{{\bar x}\}$ are the coordinates on the Euclidean manifold.
 From equations~\eqref{Noether current} and~\eqref{Noether charge
density}, we see
that~$N^{-1}t^a\mathcal{L}_\sss{\text{G}}=-\sqrt{h}\nabla_b
(n^{ab}Q[t])$ where $\rho^a$ vanishes because the solution is static.
The first integral contributes both a boundary term on the event
horizon, $\mathcal{H}$, and a boundary term term on the quasilocal
boundary~$\mathcal{B}$ that cancels the second integral.  (Note that
the normal to the event horizon is directed radially \emph{inwards} so
that~$u_an_bn^{ab}=+1$ on the event horizon.)  Also, $\int
d\tau=\Delta\tau=2\pi/\varkappa_\sss{\text{H}}$, so we find that the
entropy is
\begin{equation}
  S \approx \frac{2\pi}{\varkappa_\sss{\text{H}}} \int_{\mathcal{H}}
  \sqrt{\sigma} Q[t] d^{(n-2)}\!x \qquad \text{(zeroth order).}
  \label{entropy I}
\end{equation}

Let us evaluate equation~\eqref{entropy I} to obtain an explicit
expression for the entropy.  Recall the expression for $Q[t]$ given in
equation~\eqref{Noether Charge [t] I}.  On the event
horizon~$\mathcal{H}$, we have~$t_a=0$
and~$\nabla_at_b=-\varkappa_\sss{\text{H}}n_{ab}$.  (Notice that our
sign convention for the bi-normal $n^{ab}$ is the opposite of that
used in Iyer and Wald~\cite{IW94}.)  Thus,
$Q[t]=2\varkappa_\sss{\text{H}}D(\varPsi)$, and the entropy is:
\begin{equation}
  S \approx 4\pi \int_{\mathcal{H}} \sqrt{\sigma} D(\varPsi)
  d^{(n-2)}\!x \qquad \text{(zeroth order).}  \label{entropy II}
\end{equation}
{}From equation~\eqref{var micro action}, it is easy to calculate the
variation of the entropy amongst classical solutions.  Define
\begin{equation}
  \beta = \int N d\tau \qquad \text{and} \qquad \beta\omega^a = \int
  N^a d\tau \label{rec temp ang vel}
\end{equation}
(evaluated on the quasilocal surface) to be the reciprocal temperature
and the angular velocity of the quasilocal surface.  (Notice that
these will generally be functions over the quasilocal surface.)  Thus,
we find that
\begin{equation}
  \delta S \approx \int_{\mathcal{B}} \beta ( \delta \mathcal{E} -
  \omega^a \delta \mathcal{J}_a + \half \mathcal{S}^{ab} \delta
  \sigma_{ab} + \mathcal{Y} \delta \varPsi ) d^{(n-2)}\!x.
  \label{first law}
\end{equation}
This is the integral form of the first law of thermodynamics to zeroth
order in quantum corrections.

\subsection{Dilaton Maxwell Field Contribution}

Various non-gravitational terms will contribute additional work terms
to the first law of thermodynamics.  To see explicitly how this
occurs, we resume the study of the dilaton Maxwell action.  The
projection of the Maxwell field potential onto the
boundary~$\mathcal{T}$ is~${\frak U}_a$.  If we decompose this vector
into a piece normal to the quasilocal surface~$\mathcal{B}$, ${\frak
V}=-u^a{\frak U}_a$, and a piece projected onto the quasilocal
surface, ${\frak W}_a=\sigma^b_a{\frak U}_b$, we obtain $\delta{\frak
U}_a=N^{-1}u_a\bigl(\delta(N{\frak V})-{\frak W}_b \delta
N^b\bigr)+\sigma^b_a\delta{\frak W}_b$.  (Some of the gauge freedom of
the field potential is used to ensure that the scalar potential~$\frak
V$ remains finite even on the event horizon.)  Thus, we find that
\begin{align}
  (\varpi^a - \varpi^a_\sss0) \delta{\frak U}_a &= - \mathcal{Q}
  \delta (N{\frak V}) + (\mathcal{J}_\sss{\text{EMF}})_a \delta N^a +
  N \mathcal{K}^c \delta {\frak W}_c, \label{decomp bnd Maxwell
  momentum} \\ \intertext{where} \mathcal{Q} &=
  -\sqrt{\sigma}\,W(\varPsi) n^a{\frak E}_a - \mathcal{Q}_\sss0,
  \label{charge density}\\ (\mathcal{J}_\sss{\text{EMF}})_a &=
  \mathcal{Q}{\frak W}_a - (\mathcal{J}_\sss{\text{EMF,0}})_a,
  \label{EMF density}\\ \intertext{and} \mathcal{K}^a &=
  -\sqrt{\sigma}\,W(\varPsi){\frak F}^{cd}\sigma^a_c n_d -
  (\mathcal{K}_\sss0)^a \label{current density}
\end{align}
are the quasilocal surface Maxwell charge density, electro-motive
force (\textsc{emf}) density, and current density respectively.  They
are obtained from analogous decompositions of the
momenta~$\varpi^a-\varpi^a_\sss0 =(\delta I^1/\delta{\frak
U}_a)_{c\ell}$.  Notice that the contributions of a reference
spacetime have been included.  The Maxwell charge density and
electro-motive force are extensive variables as is the projection,
${\frak W}_a$, of the field potential onto the quasilocal surface.

The quasilocal surface Maxwell charge density can be used to find a
conserved Maxwell field charge contained within the system.  Recall
the Maxwell field equation of motion, $(E_{\frak A})^a={\frak J}^a$,
where ${\frak J}^a$ is a source and $(E_{\frak A})^a$ is given by
equation~\eqref{Maxwell EOM}:
\begin{equation}
  \nabla_b \bigl( W(\varPsi) {\frak F}^{ab} \bigr) = {\frak J}^a.
  \label{Maxwell EOM w/source}
\end{equation}
The left hand side is identically divergenceless, and
thus~$\nabla_a{\frak J}^a=0$.  Therefore, the quantity
\begin{equation}
  {\frak Q} = \int_\varSigma \sqrt{h} u_a {\frak J}^a d^{(n-1)}\!x
  \label{Maxwell charge}
\end{equation}
is conserved:
\begin{equation*}
  0 = -\int_{\mathcal{M}} \sqrt{-g} \nabla_a {\frak J}^a d^n\!x =
  {\frak Q}(\varSigma_{\text{final}}) - {\frak
  Q}(\varSigma_{\text{initial}}),
\end{equation*}
where we have assumed that the source vanishes on~$\mathcal{T}$.  This
conserved Maxwell charge can be expressed on the quasilocal surface
alone (Gauss' law):
\begin{equation}
  \begin{split} {\frak Q} &= \int_\varSigma \sqrt{h} u_a \nabla_b
    \bigl( W(\varPsi) {\frak F}^{ab} \bigr) d^{(n-1)}\!x \\ &=
    \int_\varSigma \sqrt{h} \mathcal{D}_b \bigl( W(\varPsi) {\frak
    F}^{ab} \bigr) d^{(n-1)}\!x \\ &= \int_{\mathcal{B}} \mathcal{Q}
    d^{(n-2)}\!x, \\ \end{split} \label{Maxwell charge II}
\end{equation}
where we have used the equation of motion~\eqref{Maxwell EOM w/source}
in the first line, the fact that~$u_a=-N\partial_a t$ in the second
line, and equation~\eqref{charge density} in the third.

Recall that the dilaton Maxwell field provides an extra contribution
to the Noether charge given by equation~\eqref{extra Noether charge}.
When the generator of the diffeomorphisms is the vector~$t^a$, we find
that
\begin{equation}
  Q_{\text{extra}}[t] = \mathcal{Q} N {\frak V} -
  (\mathcal{J}_\sss{\text{EMF}})_a N^a \label{extra Noether charge
  [t]}
\end{equation}
on the quasilocal surface; however, the extra contribution vanishes on
an event horizon.  Therefore, the dilaton Maxwell field does not
(explicitly) contribute to the entropy although it does contribute to
the first law of thermodynamics.  This contribution is obtained by the
inclusion of the terms in equation~\eqref{decomp bnd Maxwell momentum}
and the variation of equation~\eqref{extra Noether charge [t]} in the
integrand of equation~\eqref{var micro action}.  The first law of
thermodynamics then reads
\begin{equation}
  \begin{split} \delta S &= \int_{\mathcal{B}} \beta ( \delta
    \mathcal{E} - N^a \delta {\tilde{\mathcal{J}}}_a + \half
    \mathcal{S}^{ab} \delta \sigma_{ab} + \mathcal{Y} \delta \varPsi
    \\ & \qquad + {\frak V} \delta \mathcal{Q} + \mathcal{K}^a \delta
    {\frak W}_a ) d^{(n-2)}\!x, \\ \end{split} \label{Maxwell first
    law}
\end{equation}
where ${\tilde{\mathcal{J}}}^a=\mathcal{J}^a
+(\mathcal{J}_\sss{\text{EMF}})^a$ is the effective surface momentum
density including the electro-motive force.

\section{Thermodynamics in Two Dimensions}
\label{sec:2D thermo}

In two dimensions, the analyses of the previous section are greatly
simplified due to various identities.  One significant advantage of
analyzing thermodynamics in two dimensions is that the quasilocal
``surface'' is really just a point, and so the integral form of the
first law found in the previous section reduces to a non-integral
form.  We shall summarize here the results obtained when the
restriction to two dimensions is enforced.  For illustration, the
dilaton Maxwell term will be included in the action.

\subsection{Formul\ae\ for Quasilocal Quantities}
The action that is appropriate for the fixation of the boundary field
configurations under variations is
\begin{equation}
  \begin{split} I^1 &= \int_{\mathcal{M}} ( \mathcal{L}_\sss{\text{G}}
    + \mathcal{L}_\sss{\text{M}} ) dx\,dt - I^0 \\ &\qquad -
    2\int_\varSigma \sqrt{h}\,D(\varPsi)\tr(K) dx -
    2\int_{\mathcal{T}} \sqrt{-\gamma}\,D(\varPsi)\tr(\varTheta) dt.\\
    \end{split} \label{2d action1}
\end{equation}
In two dimensions, the Maxwell field simplifies greatly since the
two-form field strength is dual to a scalar: ${\frak F}_{ab}=
f\epsilon_{ab}$ where $ f=-\epsilon^{ab}\nabla_a{\frak A}_b$.
Therefore, the field equations are~$(E_g)_{ab}=\half T_{ab}$,
$(E_\varPsi)=\half U$, and~$(E_{\frak A})^a=0$ with
\begin{align}
  (E_g)_{ab} &= g_{ab}\nabla^2 D(\varPsi) - \nabla_a\nabla_b
  D(\varPsi) \notag\\ &\qquad + H(\varPsi) \bigl(
  \nabla_a\varPsi\nabla_b\varPsi - \half g_{ab} (\nabla\varPsi)^2
  \bigl) - \half g_{ab} V(\varPsi), \label{2d metric EOM} \\ T_{ab} &=
  \half W(\varPsi) g_{ab} f^2, \label{2d Maxwell SEM source} \\
  (E_\varPsi) &= \frac{dD}{d\varPsi}\,R - \frac{dH}{d\varPsi}\,
  (\nabla\varPsi)^2 - 2H(\varPsi)\nabla^2\varPsi +
  \frac{dV}{d\varPsi}, \label{2d dilaton EOM} \\ U &=
  \frac{dW}{d\varPsi}\, f^2, \label{2d Maxwell dilaton source} \\
  \intertext{and} (E_{\frak A})^a &= \epsilon^{ab}\nabla_b \bigl(
  W(\varPsi)f \bigr).  \label{2d Maxwell EOM}
\end{align}
However, the contribution from the~$\mathcal{T}$ boundary under
arbitrary field variations of the action~$I^1$ is given by
\begin{align}
  \delta I^1 |_{\mathcal{T}} &= \int_{\mathcal{T}} \bigl( (\pi^{ab} -
  \pi_\sss0^{ab} ) \delta \gamma_{ab} + (\varPi - \varPi_\sss0) \delta
  \varPsi + (\varpi^a - \varpi^a_\sss0) \delta {\frak U}_a \bigr) dt
  \label{2d var action1} \\ \intertext{with} \pi^{ab} &=
  \sqrt{-\gamma}\,\gamma^{ab} \bigl( n^c\nabla_c D(\varPsi) \bigr),
  \label{2d metric momentum}\\ \varPi &= \sqrt{-\gamma}\,2 \bigl(
  H(\varPsi) n^c\nabla_c\varPsi - \frac{dD}{d\varPsi} n^c a_c \bigr),
  \label{2d dilaton momentum}\\ \intertext{and} \varpi^a &=
  \sqrt{-\gamma}\,W(\varPsi) f u^a.  \label{2d Maxwell momentum}
\end{align}
We have used the fact that~$\varTheta_{ab}=\gamma_{ab}\tr(\varTheta)$
and~$\tr(\varTheta)=-n^c a_c$ on the one-dimensional
boundary~$\mathcal{T}$.

The terms in equation~\eqref{2d var action1} can be simplified easily.
First, we note that $\gamma_{ab}$ is a one-by-one matrix so
$\det(\gamma_{ab})=-N^2$ and
\begin{equation*}
  \begin{split} \pi^{ab} \delta \gamma_{ab} &= \bigl( n^c\nabla_c
    D(\varPsi) \bigr) (\sqrt{-\gamma}\gamma^{ab}\delta\gamma_{ab}) \\
    &= \bigl( n^c\nabla_c D(\varPsi) \bigr) (2\delta\sqrt{-\gamma}) \\
    &= 2\bigl( n^c\nabla_c D(\varPsi) \bigr) \delta N. \\ \end{split}
\end{equation*}
The momentum conjugate to the Maxwell field can be simplified
similarly if we use~$u^a\delta{\frak U}_a=-N^{-1}\delta(N{\frak V})$.
Thus, we have
\begin{align}
  \delta I^a |_{\mathcal{T}} &= \int_{\mathcal{T}} \bigl( - E \delta N
  + N Y \delta \varPsi - {\frak Q} \delta ( N {\frak V} ) \bigr)\,dt
  \label{2d Maxwell momentum II}\\ \intertext{with} E &= -2 \bigl(
  n^c\nabla_c D(\varPsi) \bigr) - E_\sss0, \label{2d quasi energy} \\
  Y &= 2n^c \Bigl( H(\varPsi) \nabla_c \varPsi + \frac{1}{N}\,
  \frac{dD}{d\varPsi}\,\nabla_c N \Bigr) - Y_\sss0, \label{2d quasi
  dilaton force} \\ \intertext{and} {\frak Q} &= W(\varPsi) f - {\frak
  Q}_\sss0.  \label{2d quasi Maxwell charge}
\end{align}
Recall that $a_c=N^{-1}h^a_c\nabla_aN$.

The Noether charge associated with diffeomorphisms induced by the
vector field~$t^a$ is easily evaluated on~$\mathcal{T}$:
\begin{equation}
  \begin{split} Q[t] &= -W(\varPsi) f t^a {\frak A}_a + n^{ab} \bigl(
    2 t_a \nabla_b D(\varPsi) + D(\varPsi) \nabla_a t_b \bigr) \\ &=
    W(\varPsi) f N {\frak V} - 2 N n^a \nabla_a D(\varPsi) + 2
    D(\varPsi) \varTheta_{ab} u^b ( N u^a + N^a ) \\ &= {\frak Q} N
    {\frak V} + N E - 2 N D(\varPsi) \tr(\varTheta), \\ \end{split}
    \label{2d Noether charge}
\end{equation}
so that the micro-canonical action is simply
\begin{equation}
  I_{\text{m}} = I^1 + \int_{\mathcal{T}} ( NE + {\frak Q}N{\frak V} )
  dt \label{2d micro action}
\end{equation}
and the contribution to the micro-canonical action via arbitrary
variations of the $\mathcal{T}$~boundary field configurations is
\begin{equation}
  \delta I_{\text{m}} |_{\mathcal{T}} = \int_{\mathcal{T}} N ( \delta
  E + Y \delta \varPsi + {\frak V} \delta {\frak Q} ) dt.  \label{2d
  var micro action}
\end{equation}
However, on the event horizon, the Noether charge is found to be~$Q[t]
= 2 \varkappa_\sss{\text{H}} D(\varPsi)$, which gives the usual
entropy:
\begin{equation}
  S = 4\pi D(\varPsi_\sss{\text{H}}).  \label{2d entropy}
\end{equation}
The first law of thermodynamics is
\begin{equation}
  T \delta S = \delta E + Y \delta \varPsi + {\frak V}\delta{\frak Q},
  \label{2d first law}
\end{equation}
where $T=\beta^{-1}=(2\pi N/\varkappa_\sss{\text{H}})^{-1}$.  Recall
that the gauge freedom of the potential~$\frak V$ is removed by the
condition that it remains finite at the event horizon.

\subsection{Schwarzschild-Like Coordinates}

Any static two-dimensional spacetime metric can be expressed in the
form
\begin{equation}
  ds^2 = -N^2(x) dt^2 + \frac{dx^2}{N^2(x)} \label{2d Schwarzschild
  metric}
\end{equation}
in the region covered by the $\{t,x\}$ coordinate system.  In this
coordinate system, we have $n^a=(0,N)$ and~$u_a=(-N,0)$.  We find that
$n^ca_c=N'$ and~$\varkappa_\sss{\text{H}}=\half(N^2)'$, where the
prime indicates differentiation with respect to the coordinate~$x$.
The quasilocal boundary is naturally placed at a value of
constant~$x_\sss{\text{B}}$; the system has an event horizon
where~$N^2(x_\sss{\text{H}})=0$ (for the largest~$x_\sss{\text{H}} <
x_\sss{\text{B}}$ when there are multiple horizons).  A curvature
singularity exists when the Ricci scalar, $R=-(N^2)''$, diverges.

{}From equations~\eqref{2d quasi energy} and~\eqref{2d quasi dilaton
force}, we obtain explicit expressions for the quasilocal energy and
dilaton force.  These are
\begin{align}
  E &= -2 N \frac{dD}{d\varPsi} \varPsi' - E_\sss0 \label{2d Schwarz
  quasi energy}\\ \intertext{and} Y &= \Bigl( 2NH(\varPsi)\varPsi' +
  2\frac{dD}{d\varPsi}N' \Bigr) - Y_\sss0.  \label{2d Schwarz quasi
  dilaton force}
\end{align}
The entropy is given by equation~\eqref{2d entropy} while the
temperature is
\begin{equation}
  T = \frac{1}{N}\,\frac{(N^2)'}{4\pi}.  \label{2d Schwarz temp}
\end{equation}
If the solution to the Maxwell equation of motion is given in terms
of~$f$, then we can use the relation~$f={\frak A}'_t$ and~${\frak
V}=-{\frak A}_t/N$ to get
\begin{equation}
  {\frak V} = -\frac{1}{N} \int_{x_\sss{\text{H}}}^{x_\sss{\text{B}}}
  f(x)\,dx, \label{2d scalar potential}
\end{equation}
where the constant of integration is set by the requirement
that~$\frak V$ be regular at~$x_\sss{\text{H}}$.  The quasilocal
Maxwell charge is still given by equation~\eqref{2d quasi Maxwell
charge}

\section{Examples}
\label{sec:examples}

In section~\ref{sec:thermo vars}, we obtained expressions for the
relevant thermodynamic variables of a system on a finite boundary, and
we obtained an integral relation between these variables that we
called the first law of thermodynamics.  In section~\ref{sec:2D
thermo}, we found these same quantities for the specific case of a
two-dimensional spacetime.  Here, the first law of thermodynamics
appears in a non-integral form because the quasilocal thermodynamic
variables are the quasilocal densities since the quasilocal surface is
a point.  Thus, the analysis of two-dimensional systems is greatly
simplified.  In this section we shall take two different
two-dimensional theories that admit black hole solutions, and show
that the definitions of the thermodynamic variables given above are
indeed consistent with the first law of thermodynamics.

\subsection{String-Inspired Theory}
Here we consider the specific form of the action \cite{HS92}:
\begin{equation}
  I = \int_{\mathcal{M}} \sqrt{-g}\,\frac{1}{2\kappa}\,\e^{-2\varPsi}
  \bigl( R + 4(\nabla\varPsi)^2 - \fourth{\frak F}^{ab}{\frak F}_{ab}
  + a^2 \bigr) d^2\!x,
\end{equation}
where $a$ is a positive constant and $\kappa$ is the coupling constant
(which depends on the gravitational constant).  This is a specific
case of our general action with an electromagnetic field with
\begin{gather*}
  D(\varPsi)=\frac{1}{2\kappa}\,\e^{-2\varPsi}, \quad
  H(\varPsi)=\frac{2}{\kappa}\,\e^{-2\varPsi}, \quad
  V(\varPsi)=\frac{a^2}{2\kappa}\,\e^{-2\varPsi}, \\ \intertext{and}
  W(\varPsi)=-\frac{1}{2\kappa}\,\e^{-2\varPsi}.
\end{gather*}
The equations of motion are given by equations~\eqref{2d metric
EOM}--\eqref{2d Maxwell EOM}.  For this action, they are
\begin{gather}
  2\nabla_a\nabla_b\varPsi - 2g_{ab}\nabla^2\varPsi + 2g_{ab}
  (\nabla\varPsi)^2 + \half g_{ab}a^2 + \fourth g_{ab} f^2 = 0,
  \label{string metric EOM}\\ R - 4(\nabla\varPsi)^2 +
  4\nabla^2\varPsi + a^2 + \half f^2 = 0, \label{string dilaton EOM}\\
  \intertext{and} \nabla_a\log f = \nabla_a\varPsi.  \label{string
  Maxwell EOM}
\end{gather}

We wish our quasilocal region to be empty with the exception of a
black hole.  Thus, we look for a electrovacuum solution possessing an
event horizon.  One such solution is the following:
\begin{align}
  f &= q\e^{2\varPsi} \label{string Maxwell field}\\ N^2 &= 1 -
  \frac{2m}{a}\,\e^{2\varPsi} + \frac{q^2}{2a^2}\, \e^{4\varPsi}
  \label{string metric} \\ \varPsi &= - \half ax \label{string
  dilaton}
\end{align}
in the Schwarzschild-like coordinates of equation~\eqref{2d
Schwarzschild metric}.  In this solution, $q$ and~$m$ are constants of
integration, while a third constant of integration, $x_\sss0$, was
absorbed into the definition of the origin.  Notice that the dilaton
field is proportional to the Schwarzschild-like coordinate~$x$ so we
will view the various quantities as functions of the dilaton rather
than functions of the coordinate.

The solution admits inner and outer event horizons given by
$\exp(-2\varPsi_\pm)=(1\pm\varDelta)m/a$,
where~$\varDelta^2=1-q^2/2m^2$.  We restrict our attention to the case
for which~$0<\varDelta\leq1$, that is, $2m^2>q^2\geq0$.  (When~$q=0$
there is only a single horizon.)  The outer event horizon is the
larger value, $\Psi_\sss{\text{H}}=\Psi_+$.  A calculation of the
Ricci scalar shows that the solution is singular
for~$\varPsi\to\infty$ ($x\to-\infty$) and is asymptotically flat
as~$\varPsi\to-\infty$ ($x\to\infty$).  A natural choice for the
reference spacetime is the solution with~$m=q=0$ for which~$N=1$
and~$f=0$ everywhere.  There is no longer an event horizon, and
spacetime is flat.

The quasilocal quantities can now be computed.  The quasilocal
boundary, $x_\sss{\text{B}}$, is chosen to be somewhere beyond the
outer event horizon.  At this point, the quasilocal energy, quasilocal
dilaton force, and quasilocal Maxwell charge are given by
equations~\eqref{2d Schwarz quasi energy}, \eqref{2d Schwarz quasi
dilaton force}, and~\eqref{2d quasi Maxwell charge}.  They are,
respectively,
\begin{align}
  E &= \frac{a}{\kappa}\,\e^{-2\varPsi}(1-N), \label{string quasi
  energy}\\ Y &= \frac{2a}{\kappa}\,\e^{-2\varPsi}(1-N) -
  \frac{2}{\kappa N}\, \Bigl( m - \frac{q^2}{2a}\,\e^{2\varPsi}
  \Bigr), \label{string quasi dilaton force}\\ \intertext{and} {\frak
  Q} &= -\frac{q}{2\kappa}, \label{string quasi Maxwell charge}
\end{align}
where we have used the flat spacetime as our reference.  With this
reference, the quasilocal energy is positive definite outside the
outer event horizon.  (The above functions are to be evaluated
at~$\varPsi=\varPsi(x_\sss{\text{B}})$.)  Since the dilaton field is
proportional to the spatial coordinate, we think of the dilaton force
as a sort of ``pressure'' associated with the ``size'' of the system.
We note that the constant of integration arising from the Maxwell
field equation of motion is directly related to the Maxwell charge.
Also, in the asymptotically flat regime (${\mit\Psi}\to-\infty$), the
quasilocal energy approaches the value~$m/\kappa$, so the quasilocal
energy as well as the mass are directly proportional to the constant
of integration arising from the equation of motion for the metric.

The entropy of the solution is evaluated using equation~\eqref{2d
entropy}.  We find that
\begin{equation}
  S = \frac{2\pi m}{\kappa a}\,(1+\varDelta), \label{string entropy}
\end{equation}
which allows us to eliminate the constant of integration, $m$, in
terms of the thermodynamic variables~$S$ and~$\frak Q$:
\begin{equation}
  m = \frac{2\pi\kappa{\frak Q}^2}{aS} + \frac{\kappa a S}{4\pi}.
  \label{string mass parameter}
\end{equation}
(Notice that $q$ can always be replaced with $\frak Q$ via
equation~\eqref{string quasi Maxwell charge}.)  The lapse, $N$, can
thus be expressed in terms of the entropy, $S$, the quasilocal Maxwell
charge, $\frak Q$, and the dilaton field, $\varPsi$, of the system
boundary.  Therefore, these form a complete set of independent
thermodynamic variables.  Note that the entropy is always positive.

The temperature at the system boundary can be computed if we use
equation~\eqref{2d Schwarz temp}; we obtain
\begin{equation}
  T = \frac{a}{2\pi N}\,\Bigl( \frac{1}{1+1/\varDelta} \Bigr) =
  \frac{1}{N}\,\Bigl( \frac{a}{4\pi} - \frac{2\pi{\frak Q}^2}{aS^2}
  \Bigr).  \label{string temp}
\end{equation}
It is clear that the temperature is always positive outside the outer
event horizon.  From equation~\eqref{2d scalar potential} we find that
the Maxwell field scalar potential is
\begin{equation}
  {\frak V} = \frac{q}{aN}\,( \e^{-ax_\sss{\text{B}}} -
  \e^{-ax_\sss{\text{H}}} ) = -\frac{2\kappa{\frak Q}}{aN} \Bigl(
  \e^{2\varPsi} - \frac{2\pi}{\kappa S} \Bigr).
\end{equation}

All the above formul\ae{} we derived by using the boundary properties
of the action; a first law of thermodynamics involving the quasilocal
quantities follows from a statistical treatment of the ensemble of
systems.  However, a more traditional thermodynamic approach can be
adopted.  Here, we \emph{assume} that the first law of thermodynamics
holds (or else it is not a thermodynamic system).  Given an expression
for the thermodynamic internal energy, we find the intensive
thermodynamic variables (temperature, dilaton force, and Maxwell
scalar potential) in terms of the extensive thermodynamic variables
(entropy, dilaton field, and Maxwell charge).  We shall adopt this
viewpoint in what follows in order to show that it is consistent with
the statistical treatment.

Let us take the thermodynamic internal energy to be the quasilocal
energy of equation~\eqref{string quasi energy}; then, the differential
of this expression can be written as
\begin{equation*}
  dE = \pfrac{\partial E}{\partial S}_{\varPsi,{\frak Q}} dS +
  \pfrac{\partial E}{\partial\varPsi}_{S,{\frak Q}} d\varPsi +
  \pfrac{\partial E}{\partial{\frak Q}}_{S,\varPsi} d{\frak Q}.
\end{equation*}
However, this has exactly the same form as an (assumed) first law of
thermodynamics with the identifications:
\begin{equation}
  T = \pfrac{\partial E}{\partial S}_{\varPsi,{\frak Q}}, \quad Y =
  -\pfrac{\partial E}{\partial\varPsi}_{S,{\frak Q}}, \quad \text{and}
  \quad {\frak V} = -\pfrac{\partial E}{\partial{\frak
  Q}}_{S,\varPsi}.  \label{string thermo vars}
\end{equation}
The evaluation of the above partial derivatives yields the same
expressions for the temperature, dilaton force, and Maxwell scalar
potential as were obtained earlier.

We now calculate a final thermodynamic quantity: the heat capacity.
Consider a process for which the Maxwell and dilaton fields are held
constant, that is, an isochoric process.  The heat capacity is then
\begin{equation}
  \begin{split} C_{\varPsi,{\frak Q}} &= \pfrac{\partial E}{\partial
    T}_{\varPsi,{\frak Q}} = T \biggl( \frac{\partial^2 E}{\partial
    S^2} \bigg|_{\varPsi,{\frak Q}} \biggr)^{-1} \\ &= \frac{NT}{T^2
    \e^{2\varPsi} + 4\pi{\frak Q}^2/aS^3}. \\ \end{split}
    \label{string heat capacity}
\end{equation}
Outside the event horizon, the temperature, entropy, and lapse are all
positive definite, so the heat capacity must also be positive
definite.

\subsection{Liouville Black Hole}

Another important two-dimensional theory of gravity is the
``$R=\kappa\tr(T)$'' theory~\cite{M91}.  In this theory, the
gravitational part of the action is
\begin{equation}
  I_\sss{\text{G}} = \int_{\mathcal{M}} \sqrt{-g}\,\frac{1}{2\kappa}\,
  \bigl( \varPsi R + \half(\nabla\varPsi)^2 \bigr) d^2\!x.
  \label{R=kT action}
\end{equation}
When a matter action is also present, the field equations are
\begin{gather}
  \nabla^2\varPsi - R = 0 \label{R=kT dilaton EOM} \\ \intertext{and}
  \half \bigl( \nabla_a\varPsi\nabla_b\varPsi - \half g_{ab}
  (\nabla\varPsi)^2 \bigr) + g_{ab}\nabla^2\varPsi - \nabla_a\nabla_b
  \varPsi = \kappa T_{ab}, \label{R=kT metric EOM}
\end{gather}
where $T_{ab}$ is the stress tensor of the matter---defined by
equation~\eqref{SEM source}.  We assume that the matter fields do not
couple to the dilaton (which implies that the stress tensor is
divergenceless when the equations of motion hold).  Inserting the
trace of equation~\eqref{R=kT metric EOM} into equation~\eqref{R=kT
dilaton EOM}, we obtain
\begin{equation}
  R = \kappa \tr(T),
\end{equation}
which determines the metric independently of the (classical) evolution
of~$\varPsi$; the evolution of~$\varPsi$ is determined from the
traceless part of equation~\eqref{R=kT metric EOM}.

As was done in reference~\cite{M94}, we choose a Liouville field,
$\varPhi$, as the matter so that
\begin{equation}
  I = I_\sss{\text{G}} + I_\sss{\text{L}} = I_\sss{\text{G}} +
  \int_{\mathcal{M}} \sqrt{-g}\, \bigl( \varLambda\e^{-2a\varPhi} -
  b(\nabla\varPhi)^2 - c\varPhi R \bigr) d^2\!x.  \label{Liouville
  action}
\end{equation}
Notice that the Liouville action, $I_\sss{\text{L}}$, contains a
non-minimal coupling term, namely, $-c\varPhi R$.  The presence of
this term raises some concern as our treatment of boundary terms has
assumed that there are no derivatives of the metric in the matter
action.  However, we can easily accommodate the Liouville field by
noticing that the Liouville action has the same form as the
(matterless) action of equation~\eqref{action} with the Liouville
field playing the r\^ole of the dilaton.  Thus, allowing for a
``second dilaton'' contribution to all of our analyses will fully
accommodate the Liouville field.  Therefore, the dilaton functions are
\begin{gather*}
  D(\varPsi) = \frac{1}{2\kappa}\,\varPsi, \qquad H(\varPsi) =
  \frac{1}{4\kappa},\\ {\hat D}(\varPhi) = -c\varPhi, \qquad {\hat
  H}(\varPhi) = -b, \quad \text{and} \quad {\hat V}(\varPhi) =
  \varLambda \e^{-2a\varPhi}.
\end{gather*}
The hatted functions refer to the ``second dilaton,'' i.e., the
Liouville field.

The Liouville field produces a stress tensor as well as its own
equation of motion:
\begin{gather}
  T_{ab} = g_{ab}\varLambda\e^{-2a\varPhi} + 2b \bigl( \nabla_a\varPhi
  \nabla_b\varPhi - \half g_{ab}(\nabla\varPhi)^2 \bigr) +
  2c(g_{ab}\nabla^2\varPhi - \nabla_a\nabla_b\varPhi) \label{Liouville
  stress} \\ \intertext{and} 2a\varLambda\e^{-2a\varPhi} -
  2b\nabla^2\varPhi + cR = 0 \label{Liouville EOM}
\end{gather}
respectively.  For this example, we restrict our interest to the case
in which $\kappa$ is positive and $\varLambda$ is negative.  In
addition, we require~$a$ positive, $b=a^2/\kappa$, and~$c>a/\kappa$.
These theoretical values will give us a reasonably attractive solution
(in terms of the thermodynamic quantities).  For convenience, we
define $d=c/a-1/\kappa$, which is also positive,
and~$\lambda^2=-\varLambda/2d$.

A solution to the field equations is
\begin{align}
  N^2 &= 1 - \frac{\lambda^2}{m^2}\,\e^{-2a\varPhi} \label{Liouville
  metric}\\ \varPsi &= 2a(\varPhi-\varPhi_\sss0) + \varPsi_\sss0
  \label{Liouville dilaton field}\\ \varPhi &= \frac{m}{a}(x-x_\sss0)
  = \frac{m}{a}\,x + \varPhi_\sss0; \qquad \varPhi_\sss0 =
  -\frac{m}{a}\,x_\sss0 \label{Liouville Liouville field}
\end{align}
in the Schwarzschild-like coordinates.  Here, $m$, $\varPsi_\sss0$,
and~$\varPhi_\sss0$ are constants of integration; we take~$m$ to be
positive.  The Liouville field is proportional to the spatial
coordinate so we will often use it \emph{as} the coordinate.

A calculation of the Ricci scalar yields the
value~$R=(2\lambda)^2\e^{-2a\varPhi}$, so that the solution is
singular for~$\varPhi\to-\infty$ ($x\to-\infty$), asymptotically flat
as~$\varPhi\to\infty$ ($x\to\infty$), and finite in the $m\to0$ limit.
The solution also possesses a single event horizon at
$\varPhi_\sss{\text{H}}=a^{-1}\log(\lambda/m)$.  The quasilocal system
is assumed to have a value of~$\varPhi$ greater
than~$\varPhi_\sss{\text{H}}$.

The quasilocal energy is
\begin{equation}
  E = -2N \bigl( D(\varPsi) + {\hat D}(\varPhi) \bigr)' - E_\sss0 =
  2Nmd - E_\sss0 \label{Liouville quasi energy}
\end{equation}
(cf.\@ equation~\eqref{2d Schwarz quasi energy}).  The dilaton force
is given by equation~\eqref{2d Schwarz quasi dilaton force}:
\begin{equation}
  Y = \frac{m}{\kappa N} - Y_\sss0.  \label{Liouville quasi dilaton
  force}
\end{equation}
There is also a Liouville force,
\begin{equation}
  \begin{split} F &= \Bigl( 2N{\hat H}(\varPhi)\varPhi' +
    2\frac{d{\hat D}}{d\varPhi}\,N' \Bigr) - F_\sss0 \\ &=
    \frac{2ma}{N} ( N^2 d - c/a ) - F_\sss0, \\ \end{split}
    \label{Liouville quasi Liouville force}
\end{equation}
which found from the analog of equation~\eqref{2d Schwarz quasi
dilaton force} for the Liouville field.

The surface gravity at the event horizon
is~$\varkappa_\sss{\text{H}}=m$; this yields, at the system boundary,
a temperature
\begin{equation}
  T=\frac{m}{2\pi N}.  \label{Liouville temp}
\end{equation}
The entropy is given by equation~\eqref{2d entropy}:
\begin{equation}
  \begin{split} S &= 4\pi \bigl( D(\varPsi_\sss{\text{H}}) + {\hat
    D}(\varPhi_\sss{\text{H}}) \bigr) \\ &= (4\pi d) \log(m/\lambda) +
    S_\sss0 \quad \text{with} \quad S_\sss0 =
    \frac{2\pi}{\kappa}(\varPsi_\sss0-2a\varPhi_\sss0) \end{split}
    \label{Liouville entropy}
\end{equation}
Notice that the parameter~$m$ can be written in terms of the extensive
thermodynamic variables:
\begin{equation}
  m = \lambda \exp \bigl( d^{-1} ( -\varPsi/2\kappa + a\varPhi/\kappa
  + S/4\pi ) \bigr) \label{Liouville mass param}
\end{equation}
and thus the lapse is given by
\begin{equation}
  N^2 = 1 - \exp \bigl( d^{-1} ( \varPsi/\kappa - 2c\varPhi - S/2\pi )
  \bigr).
\end{equation}
Therefore, the intensive variables can all be written as functions of
the extensive variables.

Let us set $E_\sss0=0$, $Y_\sss0=0$, and~$F_\sss0=0$.  This would be
consistent with defining the reference spacetime as the~$m\to0$ limit
of our solution, but there is some difficulty with this
interpretation.  Although the curvature scalar, along with all other
terms in the action, remain finite, various quantities such as the
Liouville field will have logarithmic-like divergences in the~$m\to0$
limit of our solution.  A safer interpretation is simply the
\textit{ad hoc} choice~$I_\sss0=0$.  With this choice for the zero of
quasilocal energy, we see that, in the asymptotic
limit~${\mit\Phi}\to\infty$, the quasilocal energy approaches the
value~$2md$.  Again, the quasilocal energy and mass are directly
proportional to the constant of integration that arises from the
equation of motion for the metric.

We now show that the treatment above, in which the intensive
thermodynamic variables are obtained from a statistical point of view,
is consistent with the thermodynamic approach.  As in the string case
earlier, we assume that the thermodynamic internal energy is given by
equation~\eqref{Liouville quasi energy}.  Furthermore, we assume a
first law of thermodynamics:
\begin{equation*}
    dE = T dS -Y d\varPsi -F d\varPhi.
\end{equation*}
Since
\begin{equation*}
    dE = \pfrac{\partial E}{\partial S}_{\varPsi,\varPhi} dS
     +\pfrac{\partial E}{\partial\varPsi}_{S,\varPhi} d\varPsi
     +\pfrac{\partial E}{\partial\varPhi}_{S,\varPsi} d\varPhi,
\end{equation*}
we can identify the thermodynamic relations:
\begin{equation}
  T = \pfrac{\partial E}{\partial S}_{\varPsi,\varPhi}, \quad Y =
  -\pfrac{\partial E}{\partial\varPsi}_{S,\varPhi}, \quad \text{and}
  \quad F = -\pfrac{\partial E}{\partial\varPhi}_{S,\varPsi}.
  \label{Liouville thermo vars}
\end{equation}
As expected, the thermodynamic relations of equation~\eqref{Liouville
thermo vars} result in the same values for the intensive variables as
were obtained in equations~\eqref{Liouville temp}, \eqref{Liouville
quasi dilaton force}, and \eqref{Liouville quasi Liouville force}.

Finally we compute the heat capacity for constant dilaton and
Liouville fields:
\begin{equation}
  \begin{split} C_{\varPsi,\varPhi} &= \pfrac{\partial E}{\partial
    T}_{\varPsi,\varPhi} = T \biggl( \frac{\partial^2 E}{\partial S^2}
    \bigg|_{\varPsi,\varPhi} \biggr)^{-1} \\ &= \frac{4\pi N^2 d}{2N^2
    - 1}. \\ \end{split} \label{Liouville heat capacity}
\end{equation}
The heat capacity depends on the position of the system boundary in an
interesting way.  At a critical system size,
$\varPhi_{\text{crit}}=\varPhi_\sss{\text{H}}+a^{-1}\log2$, for
which~$N^2=\half$, the heat capacity diverges.
For~$\varPhi<\varPhi_{\text{crit}}$, the heat capacity is negative,
but approaches zero as~$\varPhi\to\varPhi_\sss{\text{H}}$.
For~$\varPhi>\varPhi_{\text{crit}}$, the heat capacity is positive and
it approaches the value~$4\pi d$ as~$\varPhi\to\infty$.
Qualitatively, this is the opposite to what we find in the usual
four-dimensional Schwarzschild solution in General Relativity.

\section{Concluding Remarks}
\label{sec:conclusions}

We have discussed the general formulation of gravitational
thermodynamics for a wide class of dilaton theories coupled to
electromagnetism in $n$ dimensions.  From this, we have shown how to
derive various thermodynamic quantities relevant to two spacetime
dimensions, and we applied this to two model black hole spacetimes in
different two-dimensional theories.  These quantities were obtained
from boundary terms of the action for which the boundary field
configurations are held fixed in deriving the field equations from a
variational principle.  Furthermore, we have obtained a relationship
amongst the variables that is the first law of thermodynamics.  In
this final section, we will speculate on the nature of the evaporation
of our two model spacetimes using thermodynamic techniques.

Our na{\"\i}ve approach is based on the black body radiation law in
one spatial dimension:
\begin{equation}
  F=\sigma T^2, \label{black body law}
\end{equation}
where $F$ is the flux emitted out of the quasilocal region and
$\sigma=\pi/12$ is the Stefan-Boltzmann constant.  For one spatial
dimension, the flux is just the negative of the time rate of change of
the thermodynamic internal energy (the quasilocal energy) of the
quasilocal region.  If the temperature of the reservoir surrounding
the quasilocal region is less than the temperature on the boundary of
the quasilocal region, then energy will be lost from the system
according to equation~\eqref{black body law}, and the black hole
contained will evaporate.

Let us first consider the string-like black hole solution.  Let us
also restrict our attention to a quasilocal boundary in the
asymptotically flat regime, $\varPsi\to-\infty$.  In this limit, we
have $E=m/\kappa$ and~$T=(a/2\pi)(1+1/\varDelta)^{-1}$.  (Recall
that~$\varDelta^2=1-q^2/2m^2$.)  As energy is lost from the quasilocal
region, the mass parameter~$m$ shrinks according to
\begin{equation}
  \frac{dm}{dt} = -\frac{\kappa\sigma a^2}{2\pi}\,
  \pfrac{1}{1+1/\varDelta}^2.  \label{string evap de}
\end{equation}
It is simple to show that the length of time, $t_{\text{evap}}$,
required for the black hole to evaporate to the extremal state, in
which the mass is~$q/\sqrt{2}$ and~$\varDelta=0$, diverges.  This is
fortunate since otherwise the system could radiate to zero temperature
in a finite time, and it would thus violate the third law of
thermodynamics.  However, the uncharged black hole evaporates
completely (i.e., to zero mass parameter) in time
\begin{equation}
  t_{\text{evap}}=\frac{16\pi^2}{\sigma\kappa a^2}\,m.
  \label{uncharged string evap time}
\end{equation}
In this qualitative respect then, the charged string-like black hole
solution behaves like the Reissner-Norstr\o m solution, and the
uncharged string-like black hole behaves like the Schwarzschild
solution. Note that this does not violate the third law of
thermodynamics as the temperature remains at the constant value
of~$a/4\pi$ (a feature that is qualitatively different from the
Reissner-Norstr\o m and Schwarzschild solutions).

Now let us discuss the Liouville black hole solution.  At a fixed
radius, $\varPhi_\sss{\text{B}}$, the quasilocal energy and
temperature are~$E=2Nmd$ and~$T=m/(2\pi N)$ respectively, where
$N^2=1-\lambda^2\exp(-2a\varPsi_\sss{\text{B}})/m^2$.  Thus, the mass
parameter of the black hole will decrease according to
\begin{equation}
  \frac{N}{m^2}\,dm = -\frac{\sigma}{8\pi^2 d}\,dt.  \label{Liouville
  evap de}
\end{equation}
(We hold the coordinate position of the quasilocal boundary, and
thus~$\varPhi_\sss{\text{B}}$, fixed during the evaporation.)  As the
mass parameter decreases, the black hole expands until eventually the
event horizon hits the boundary of the quasilocal region when the mass
parameter has fallen to the value
of~$\lambda\exp(-a\varPhi_\sss{\text{B}})$.  The time necessary to
achieve this is
\begin{equation}
  t_{\text{evap}} = \frac{4\pi^2 d}{\sigma} m \biggl(
  \frac{\arccos\bigl(\sqrt{1-N}\bigr)}{\sqrt{1-N}} - N \biggr).
  \label{Liouville evap time}
\end{equation}
At this time, the temperature of the system has become infinite, so
again the third law of thermodynamics is not violated.

\paragraph{Acknowledgments}
This work was supported in part by the Natural Sciences and
Engineering Research Council of Canada.

\end{document}